\documentstyle[11pt,paspconf,epsf]{article}
\begin{document}

\title{Using weak Mg\,II lines to chart Low Surface Brightness Galaxies}
\author{V. Le Brun}
\affil{IGRAP/LAS du CNRS, BP 8, 13376 Marseille CEDEX 12}
\author{C.W. Churchill}
\affil{Penn. State University, University Park, PA 16802}
\begin{abstract}
We report the detection, based on HST and Keck data, of two peculiar 
absorbers in the Ly$\alpha$ forest of the quasar PKS~0454+039. These clouds, 
at redshifts $z=0.6248$ and $0.9315$ respectively, display both Mg\,{\sc ii} and 
Fe\,{\sc ii} absorption lines in addition to the Ly$\alpha$ line. Based upon
photoionization models, these are inferred to be photoionized by the
intergalactic UV background, and to have H\,{\sc i} column densities in the
range $15.8\le N(\mbox{H\,{\sc i}}) \le 16.8$. Furthermore, if one supposes that the relative
abundances of heavy elements is similar to that of depleted clouds of our
galaxy, the abundances of these two absorbers are greater than the solar
value, which is a unique case for absorbers which are not associated to the
quasar. We tentatively suggest that these absorbers may select giant low
surface brightness galaxies.
\end{abstract}


\section{Introduction}
Quasars, beside the interest of their internal mechanisms, are unique tools
for the study of the gaseous content of the Universe. In fact, any single
gaseous cloud (down to a surface density of a few $10^{12}$~cm$^{-2}$) placed 
on the sightline to one of these bright and distant point sources can
leave an imprit, in the form of one or several absorption lines, on the
spectrum of the latter.

Absorption line systems (a system is a set of several lines of different
ionization states of different elements) are classified following their
content in neutral hydrogen :
\begin{itemize}
\item The so-called Damped Ly$\alpha$ systems (DLAS), because their Ly$\alpha$
line is located on the damped part of the curve of growth, are characteristic
of large neutral hydrogen column densities ($N(\mbox{H\,{\sc i}}) \ge
2~10^{21}$~cm$^{-2}$). Their similarity with Galactic clouds in term of
physical conditions (ionization degree, temperature ...), led people to
associate them to distant spiral galaxies (Wolfe et al. 1986). In fact, they 
where recently confirmed, thanks to HST imaging and spectroscopy, to be 
associated with galaxies of any type (Le Brun et al. 1997),
\item The Mg\,{\sc ii} systems display less neutral hydrogen than DLAS, even if they
are still optically thick ($N(\mbox{H\,{\sc i}})\ge 10^{17}$~cm$^{-2}$). The gas of these
clouds is of low ionization level, and they have been shown to be associated
with large ($R\sim 90h_{50}^{-1}$~kpc) gaseous halos of bright field galaxies
(Bergeron \& Boiss\'e 1991, Steidel 1995).
\item At last, the absorbers of the Ly$\alpha$ forest outnumber the other
classes by several decades, each system displaying only the Ly$\alpha$ line
(at least with an average spectroscopic set-up), $N(\mbox{H\,{\sc i}})$ being in the range
$10^{12} - 10^{16.5}$. Photoionization models show that the gas of these
clouds is highly ionized, with some of them displaying ionic lines from
C\,{\sc iv}, N\,{\sc v} or O\,{\sc vi}. The abundances are quite low : $Z\simeq
10^{-3} - 10^{-2} Z\odot$. These clouds could trace both the external parts of
the galactic halos and the intergalactic gas spread along the large scale
structure of the galaxy distribution (Le Brun \& Bergeron 1998). 
\end{itemize}
In this paper, we present the complete study of two absorbers of a new kind,
the ``weak Mg\,{\sc ii} absorbers''. Section~2 presents the state-of-art about these
absorbers and Sect.~3 the peculiar sightline to the quasar PKS~0454+039, with 
all data we obtained on it. At last, Sect.~4 presents our discussion and conclusions 
about these objects.

\section{The weak Mg\,II absorbers}
\begin{figure}
\plotone{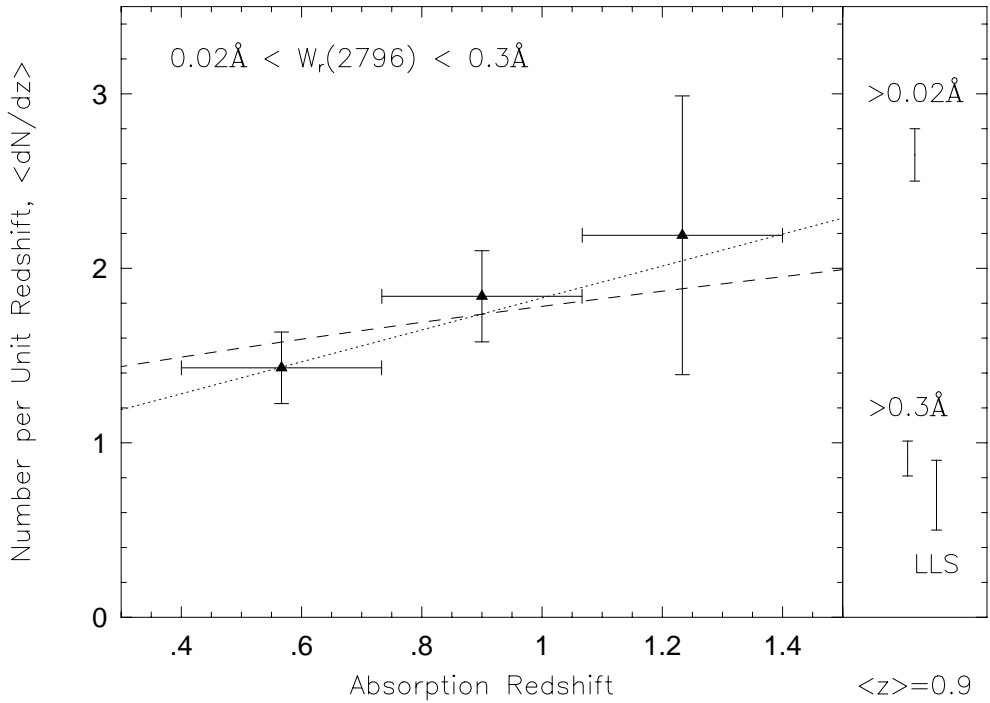}
\caption{\label{dndz} Redshift evolution of the weak Mg\,{\sc ii} absorbers
  population.} 
\end{figure}

The Mg\,{\sc ii} systems were extensively studied in the 80's thanks to systematic
spectroscopic surveys of several tens of quasars at signal to noise ratio
about 10, with limiting rest equivalent width of about 0.3~\AA. The most complete 
of those was published in 1992 (Steidel \&
Sargent). These surveys have shown that the number density, either in value or
in evolution with redshift, was fully compatible with these absorbers being
linked with field galaxies. This hypothesis was confirmed in the same time by
the the first identifications of absorbing galaxies (Bergeron \& Boiss\'e
1991). It is only with the advent of the
Keck Telescope that Churchill et al. (1998) could initiate a survey for
``weak'' Mg\,{\sc ii} absorbers, that is with rest equivalent width down to
0.02~\AA. The survey was made with HIRES (Vogt et al. 1994), all details are
given in Churchill et al. (1998).
\begin{figure}
\plotone{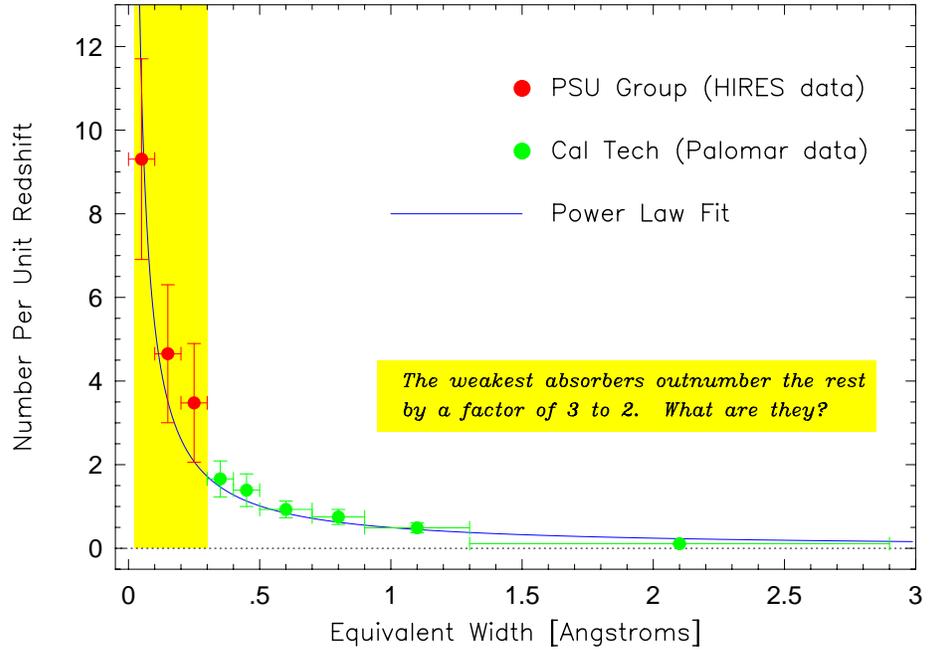}
\caption{\label{nw} Rest equivalent width distribution of the weak Mg\,{\sc ii}
  absorbers (in the shadened zone) as compared to strong ones (on the right of 
the diagram)}
\end{figure}
 As can be seen on Fig~\ref{nw}, the weak
Mg\,{\sc ii} absorber outnumber the strong ones by a factor of 2 to 3, and there is
no lower cutoff in the distribution of the rest equivalent width down to
0.02~\AA. Also, the evolution of the number density of weak absorbers show
that it is compatible with a non evolving (in number) population (Fig.~\ref{dndz}). 
\begin{figure}
\plottwo{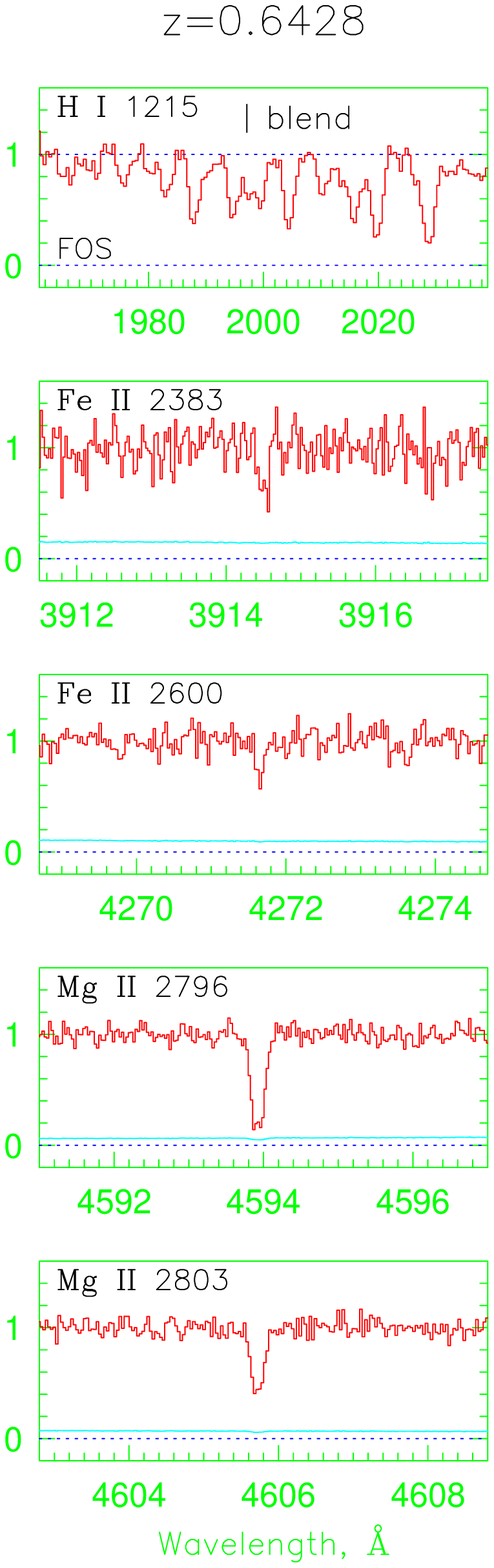}{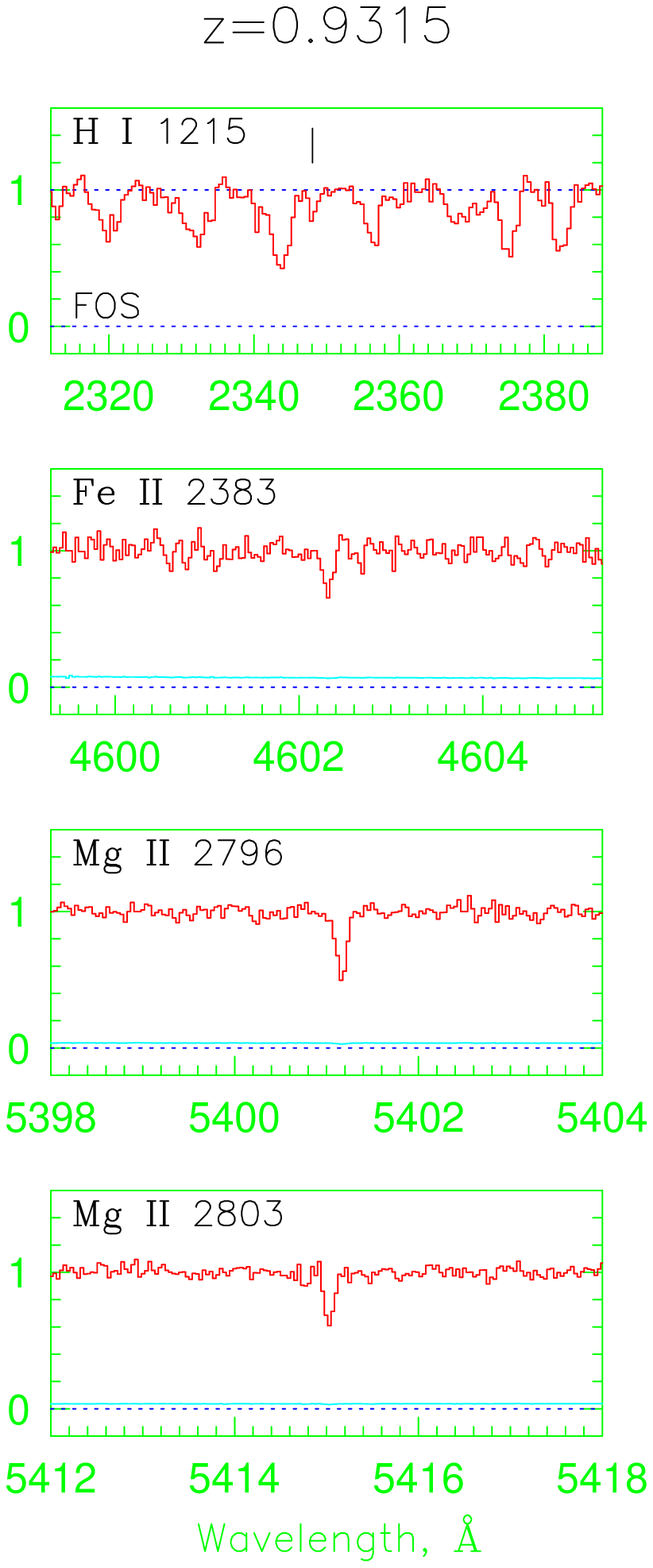}
\caption{\label{spectra}FOS/HST (top panel) and Keck/HIRES (lower panels) 
    spectra of the two absorbers. Note the very different wavelength scales 
    between the optical and UV spectra}
\end{figure} 
If these absorbers are of the same nature as the strong ones, by comparing
the density of absorbers and of galaxies, it requires galaxies to be
surrounded by halos of radius $R\sim120h_{50}^{-1}$~kpc. However, this value
can be lowered if a fraction of the weak absorbers is perhaps of different origin
for example Low Surface Brightness (LSB) galaxies. 
\section{The sightline to PKS 0454+039 : data and analysis}
To ascertain the nature of weak Mg\,{\sc ii} absorbers, we have focused on the
sightline toward the quasar PKS~0454+039. This object is part of the
Keck/HIRES survey, and a high resolution ($R\sim 45,000$) high signal to noise
ratio ($\sim 50$) has been obtained in the visible domain. Beside of this, the
HST/FOS UV spectrum has been extensively studied by Boiss\'e et al. (1998) for
the characterization of the $z=0.8596$ DLAS present in front of the quasar. The
spectral resolution is $R\simeq 1300$, and the signal to noise ratio is about
10. The limiting equivalent width for a $3\sigma$ detection is about
0.3~\AA. At last, we have obtained deep CFHT and HST/WFPC2 $R$ band images of
the field surrounding this quasar (Le Brun et al. 1997).

The analysis of the HIRES spectrum shows that two faint Mg\,{\sc ii} absorbers are
present at redshifts $z=0.6248$ and $z=0.9315$, together with Fe\,{\sc ii} lines at 
the same redshift
%. However, the strongest
%surprise what the presence of Fe\,{\sc ii} lines at the same redshifts. 
%In fact, the
%presence of Fe\,{\sc ii} at a level comparable to Mg\,{\sc ii} indicates, as derived from
%photoionization models (see e.g. Bergeron \& Stasi\'nska 1986), the presence
%of large amounts of neutral hydrogen, and thus of a DLAS, at least for the 
%strong Mg\,{\sc ii}
%absorbers. 
Fig.~\ref{spectra} displays both the Ly$\alpha$ line, as present in the
HST/FOS spectrum (top panels), and the metal lines (lower panels). As can be
seen, the Ly$\alpha$ lines are very faint, and were even not
included in the $3\sigma$ limited sample of absorption lines listed in
Boiss\'e et al. (1998). Their rest equivalent width are 0.33 and 0.15~\AA\
respectively. 

We have thereafter tried to use the standard analysis methods to derive the
physical  properties of the gas. However, since, even at the HIRES resolution,
the Mg\,{\sc ii} and Fe\,{\sc ii} lines are barely resolved, we have used Monte Carlo
simulations to determine the best values for the column densities and
dispersion parameters, using the doublet ratio. We obtain that, for both
systems, $N(\mbox{Fe\,\sc ii}) \sim N(\mbox{Mg\,\sc ii}) \sim 10^{12.5}$,
while the $b$ parameter are  5 and 2~km\,s$^{-1}$ for the $z=0.6428$ and
0.9315 systems respectively. 

Unfortunately, the FOS spectrum is of poorer quality, and a Voigt profile
fitting was impossible, so that we could not derive directly the properties of
the H\,{\sc i} gas. We have therefore proceeded in several steps that are
summarized below (see Churchill \& Le Brun 1998 for a detailed description
of this work):
\begin{figure}
\plotone{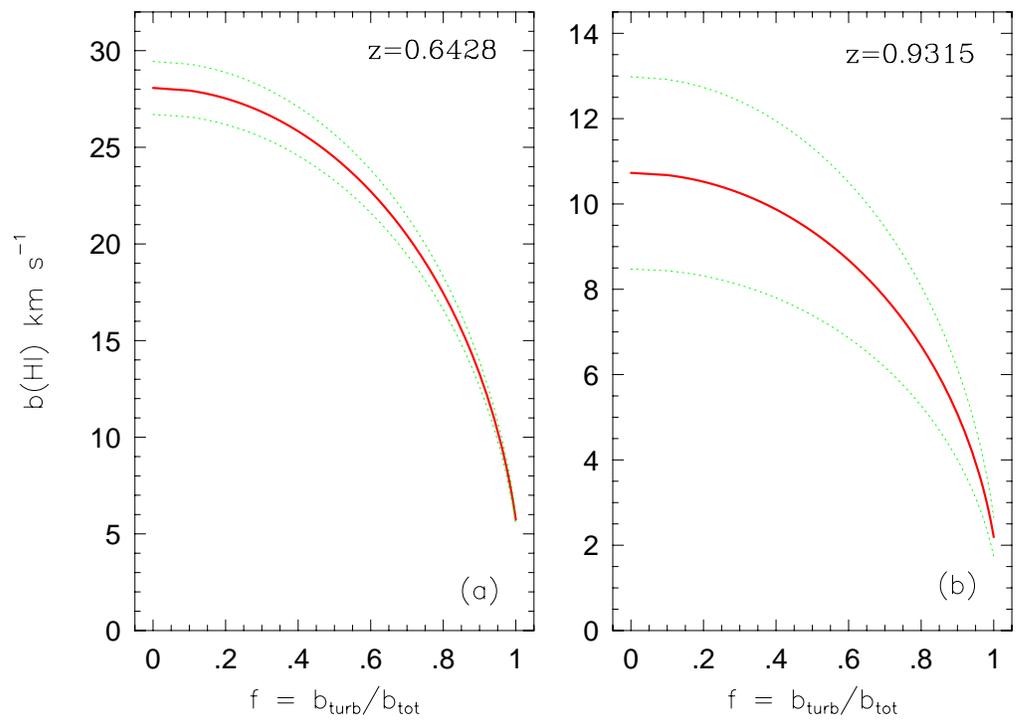}
\caption{\label{fb} Variation of the H\,{\sc i} broadening value as a function
  of the turbulence parameter for the two absorbers}
\end{figure}
\begin{enumerate}
\item We have introduced the turbulence  parameter, $f=b_{\mathrm turb}/
b_{\mathrm tot}$, which can have values between 0 and 1. If $f=0$, the gas is 
thermally excited,
and the $b$ parameter for different elements scales as the square root of the
mass ratio. On the contrary, if $f=1$, the gas is fully collisionally ionized,
and all the lines of all elements have the same $b$ value. Of course, all
intermediate situations are possible. The variation of $b(\mbox{H\,{\sc i}})$
as a function of the $f$ value is shown on Fig.~\ref{fb} for the two systems.
\item From this range of possible variations for the $b$ parameter, we thus
can derive, using the curve of growth analysis, the range of possible values
for the neutral hydrogen column densities : it covers nearly 3 decades from
$\sim 10^{14}$ to $\sim 10^{17}$~cm$^{-2}$. 
\item The latter result  makes it impossible to derive
any hints on the physical states of these absorbers just from the data. We
thus have used CLOUDY (Ferland 1996), to make some simulations of the
absorbing gas. For each value of $N(\mbox{H\,{\sc i}})$ between $10^{14}$ and
$10^{17}$, by step of 0.5 in log, we have run CLOUDY in 'optimized' mode, so
that the simulation converges toward the observed values of $N(\mbox{Fe\,{sc ii}})$
and $N(\mbox{Mg\,{\sc ii}})$. The other inputs are i) the UV ionizing external 
radiation
field : it could either have a galactic-shaped spectrum, or an intergalactic 
UV background shape, as given by Haardt \& Madau (1996), and ii) the abundance
pattern, that is the relative abundances of heavy elements : solar, H\,{\sc ii}
region, that is including depletion by dust, or enhanced abundances of
$\alpha$ elements. The output of CLOUDY
is the full physical state of the gas, including temperature, from which we
could derive the $f$ parameter value. Thereafter, we only had to compare the
output of the simulation to the observational constraints to get the possible
values of $N(\mbox{H\,{\sc i}})$.
\end{enumerate}
\section{Discussion and conclusions}
\begin{figure}
\plottwo{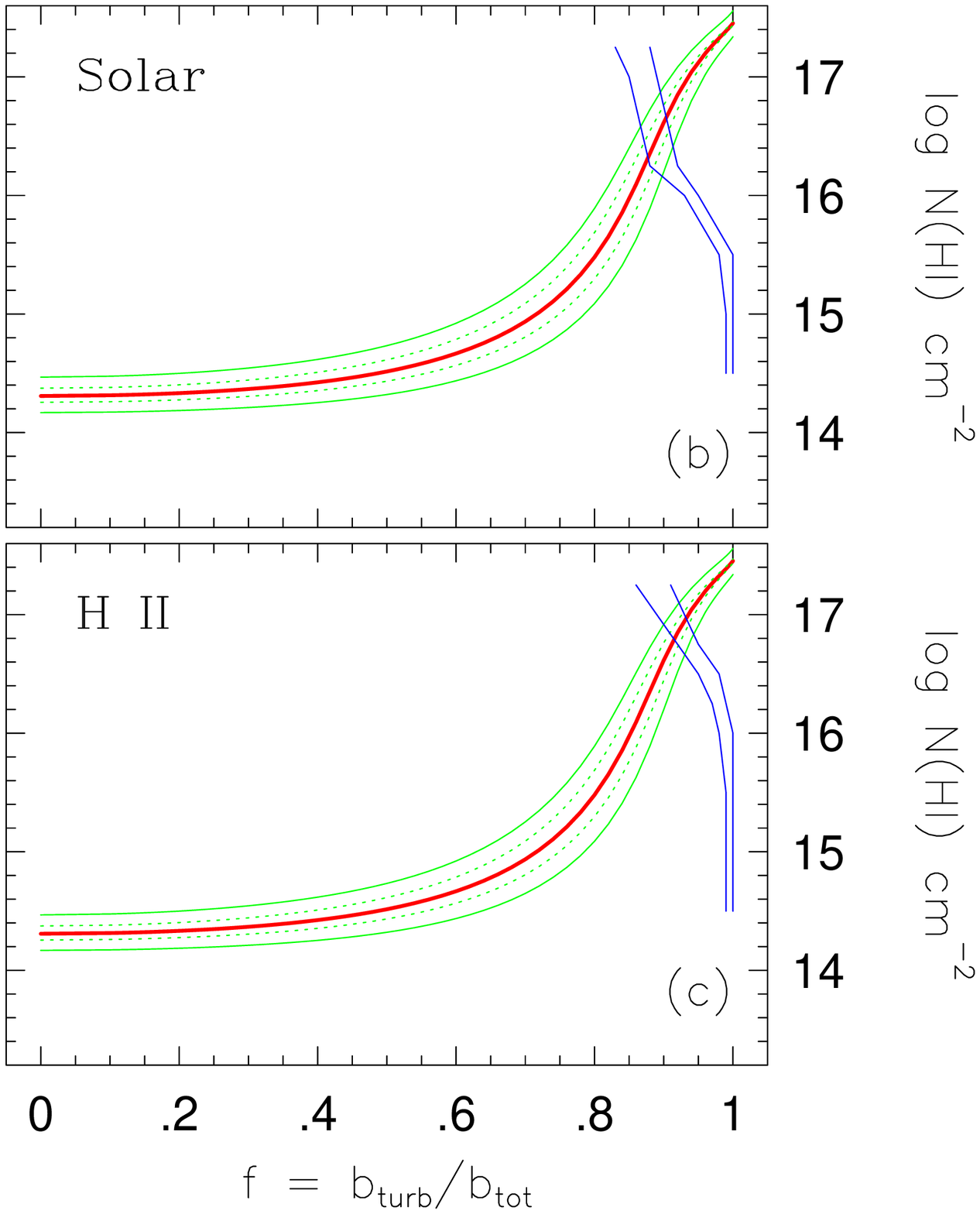}{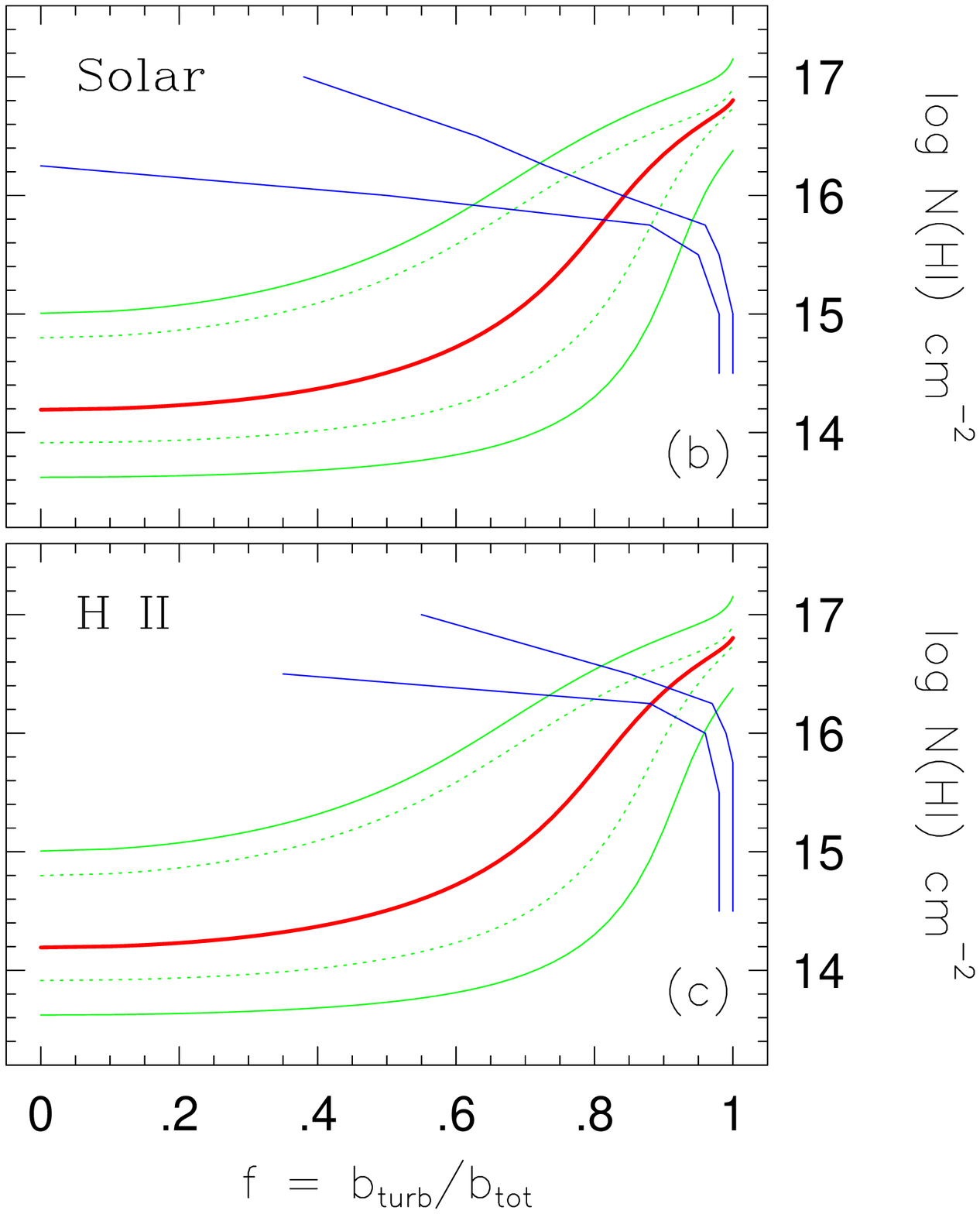}
\caption{\label{fnhi} Properties of the two absorbers ($z=0.6428$ on the left,
  $z=0.9315$ on the right). Thick solid lines give the value of 
  $N(\mbox{H\,{\sc i}})$ derived from observed $b_{\mathrm tot}
  (\mbox{Mg\,{\sc ii}})$ and $W_{\mathrm r}(\mbox{H\,{\sc i}})$. Thin curves 
  give the uncertainties from the measurements. The curves that originate in 
  the lower-right corner and rise upwards and then to the left are the allowed 
  locus of $f$ for a cloud model of a given $N(\mbox{H\,{\sc i}})$}
\end{figure}

These simulations allowed us to eliminate some hypothesis : the
$\alpha$-enhanced abundance pattern fails to produce coherent models, as well
as the galactic-shaped ionizing flux, which requires unrealistic spatial
densities of stars to reproduce the physical quantities of the gas. Thus,
Fig.~\ref{fnhi} displays the domain that is allowed in the $f -
N(\mbox{H\,{\sc i}})$ plane for the two absorbers : it is the intersection of
the band coming from lower-left to upper right, which displays the range
allowed by data (with the errors), and the band going from lower-right to
upper-right, which reflects the uncertainties in the Fe\,{\sc ii} and Mg\,{\sc ii} column
densities. As can be seen, only a small domain is allowed for each absorber,
which covers less than a decade in $N(\mbox{H\,{\sc i}})$. 

As a result of this, we can now get estimates of the metallicity of these
clouds, which are surprisingly high : the $z=0.6428$ absorber has abundances
$Z\ge 0.2 Z_\odot$ if the abundance pattern is solar, and $Z\ge 1.6 Z_\odot$ if
the gas has abundances similar to the Galactic H\,{\sc ii} regions, i.e with
depletion on dust grains. For the $z=0.9315$ absorber, the abundances are
above the solar value, whatever the abundance pattern. 

Their very high abundances make these absorbers very peculiar, and in any case
different from the strong Mg\,{\sc ii} absorbers, which have metallicities 
$Z\sim 0.01 Z_\odot$, and at least a part of the weak Mg\,{\sc ii} absorbers 
thus seems not
to originate in field galaxy halos. Furthermore, we have searched the deep
CFHT and HST images of the field around the quasar, and there is no galaxy
close enough to the sightline that  could give rise to these absorption
systems, when one takes into account the galaxies that are likely to host the
four already know ``normal'' metallic absorption systems.

There is  however a class of galaxies which present the same characteristics
as our absorbers : The Giant Low Surface Brightness Galaxies. Their
abundances, as measured in H\,{\sc ii} regions, are above the solar value
(McGaugh 1994, Pickering \& Impey 1995), and the H\,{\sc i} gas velocity
dispersion is quite low, in the range $10 - 30$~km\,s$^{-1}$, thus similar to
the value derived in our absorbers. These two similarities, which can not be
found in any other class of galaxies, leads us to suggest that a least a
fraction of the weak Mg\,{\sc ii} absorbers is due to these giant LSBGs. The spatial
densities of absorbers and galaxies cannot be compared yet, since the samples
of both are to small to derive useful statistics.

The immediate follow-up of this work should go in two direction : first, more
detailed UV spectroscopy is necessary, to derive better constraints on both the
Ly$\alpha$ profile and other ions absorption lines (C\,{\sc iv}, O\,{\sc iii},
O\,{\sc vi}, ...), and also by more imaging and spectroscopy in the field, to
try to identify the absorbing galaxies, or companions of them. These
development will need large aperture space and ground-based telescopes.

\end{document}